\documentclass[prl,nofootinbib,twocolumn,superscriptaddress]{revtex4-1}

\usepackage[utf8]{inputenc}

\usepackage[colorlinks]{hyperref}
\hypersetup{
     colorlinks   = true,
     citecolor    = red,
	linkcolor=blue
}

\usepackage{bbm}
\usepackage{amsfonts}
\usepackage{mathrsfs}
\usepackage{slashed}
\usepackage{bbold}
\usepackage{amsmath,amssymb}
\usepackage{epsfig}
\usepackage{graphicx}               
\usepackage{url}
\usepackage{float}
\usepackage{soul}
\usepackage{pstricks}
\usepackage{color}
\usepackage{multirow}

\newcommand{\LL}{\mathcal{L}}
\newcommand{\MM}{\mathcal{M}}
\newcommand{\OO}{\mathcal{O}}

\newcommand\Tstrut{\rule{0pt}{3.5ex}}         
\newcommand\Bstrut{\rule[-2ex]{0pt}{0pt}}   

\def\vect#1{\boldsymbol{#1}}

\def\newpar{\vskip4pt}

\allowdisplaybreaks

\begin{document}

\title{Detecting Light Dark Matter with Magnons}

\author{Tanner Trickle}
\author{Zhengkang Zhang}
\affiliation{Department of Physics, University of California, Berkeley, CA 94720, USA}
\affiliation{Theoretical Physics Group, Lawrence Berkeley National Laboratory, Berkeley, CA 94720, USA}
\affiliation{Walter Burke Institute for Theoretical Physics, California Institute of Technology, Pasadena, CA 91125, USA}

\author{Kathryn M.~Zurek}
\affiliation{Walter Burke Institute for Theoretical Physics, California Institute of Technology, Pasadena, CA 91125, USA}
\affiliation{Theoretical Physics Group, Lawrence Berkeley National Laboratory, Berkeley, CA 94720, USA}
\affiliation{Department of Physics, University of California, Berkeley, CA 94720, USA}

\begin{abstract}
Scattering of light dark matter with sub-eV energy deposition can be detected with collective excitations in condensed matter systems. When dark matter has spin-independent couplings to atoms or ions, it has been shown to efficiently excite phonons.  Here we show that, if dark matter couples to the electron spin, magnon excitations in materials with magnetic dipole order offer a promising detection path. We derive general formulae for single magnon excitation rates from dark matter scattering, and demonstrate as a proof of principle the projected reach of a yttrium iron garnet target for several dark matter models with spin-dependent interactions. This highlights the complementarity of various collective excitations in probing different dark matter interactions.
\end{abstract}

\maketitle


\paragraph{Introduction}---\,
Direct detection of dark matter (DM) has undergone a dramatic expansion of scope in recent years.  Well-motivated theories where sub-GeV DM arises in a hidden sector/hidden valley, with new weakly or strongly coupled dynamics (see {\it e.g.}\ Refs.~\cite{Carlson:1992fn,Boehm:2003hm,Strassler:2006im,Pospelov:2007mp,Hooper:2008im,Feng:2008ya,Lin:2011gj,Hochberg:2014dra} for early examples), have given impetus to new ideas to search for light DM. Conventional nuclear recoils, well-matched kinematically to search for weak-scale DM, are not effective for light DM --- once the DM mass drops below the target nucleus mass, the fraction of the DM's kinetic energy that can be deposited on the target falls. Beyond nuclear recoils, better DM-target kinematic matching allows us to probe qualitatively new parameter space, through lighter targets ({\it e.g.}\ electrons) with $\sim$\,eV (as in semiconductors and atoms \cite{Essig:2011nj,Graham:2012su,Essig:2012yx,Lee:2015qva,Essig:2015cda,Derenzo:2016fse,Hochberg:2016sqx,Bloch:2016sjj,Essig:2017kqs,Kurinsky:2019pgb} as well as molecules \cite{Essig:2016crl,Arvanitaki:2017nhi,Essig:2019kfe}) or $\sim$\,meV (as in superconductors \cite{Hochberg:2015pha,Hochberg:2015fth,Hochberg:2016ajh} and Dirac materials \cite{Hochberg:2016ntt}) energy gaps. Reading out such small energy depositions is achieved through improvements to cryogenic superconducting calorimeters, such as transition edge sensors (TES) and microwave kinetic inductance devices (MKIDs). {\it Collective} excitations, such as phonons in superfluid helium \cite{Schutz:2016tid,Knapen:2016cue,Acanfora:2019con,Caputo:2019cyg} and crystals \cite{Knapen:2017ekk,Griffin:2018bjn}, open new avenues for good kinematic matching. For example, the presence of $\OO(10$-$100)$\,meV gapped optical phonons in some systems facilitates the extraction of a large fraction of DM's kinetic energy for DM as light as $\sim$10\,keV.

Beyond kinematics, there is also a {\it dynamics} aspect of the problem --- depending on how the DM couples to Standard Model (SM) particles, different target responses are relevant. A familiar example from nuclear recoils is the presence of several nuclear responses -- spin-independent (SI), spin-dependent (SD), {\it etc}.\ -- which can probe different DM-SM interactions \cite{Chang:2009yt,Fitzpatrick:2010br,Fitzpatrick:2012ix,Gresham:2014vja}. Together they provide broad coverage of the DM theory space, with various target nuclei offering complementary information. Another example is dark photon mediated DM: a material with a strong optical response, such as a superconductor, has weak reach since the effective coupling of the dark photon is suppressed due to in-medium effects, while Dirac materials and polar crystals, which have weaker optical response, have excellent reach \cite{Hochberg:2016ntt,Knapen:2017ekk,Griffin:2018bjn}. Similarly, collective excitations can arise from different degrees of freedom, such as charge or spin, and some excitations may be advantageous over others for certain types of DM couplings. Therefore, in order to identify the broadest DM detection strategy, it is important to consider collective excitations of all types.

From this perspective, previous proposals via phonon excitations are aimed at probing SI responses. While they cover many simple DM models, including those with a dark photon or scalar mediator, there are other scenarios that are equally plausible, where the leading DM-SM interactions lead to stronger SD responses. For example, in dark photon mediated models, the DM may in fact be charge neutral, but couple to the dark photon via a higher multipole, {\it e.g.}\ magnetic dipole or anapole \cite{Pospelov:2000bq,Sigurdson:2004zp,Masso:2009mu,Fitzpatrick:2010br,Chang:2010en,Barger:2010gv,Banks:2010eh,Ho:2012bg,Gresham:2014vja,DelNobile:2014eta,Kavanagh:2018xeh,Chu:2018qrm}. Also, a spin-0 mediator may dominantly couple to the pseudoscalar (rather than scalar) current of SM fermions. In these scenarios, summarized in Table~\ref{tab:models}, SI responses are suppressed compared to the previously considered cases, and ideas of detecting SD responses are needed. More generally, SI and SD couplings can coexist, so it is desirable to pursue detection channels for both in order to have a more complete picture of DM interactions.

\begin{table*}
\centering
\begin{tabular}{p{105pt}p{155pt}p{130pt}p{95pt}}
\hline  
Magnetic dipole DM & 
$\LL =  \frac{g_\chi}{\Lambda_\chi}\bar\chi\sigma^{\mu\nu}\chi \,V_{\mu\nu} +g_e\bar e\gamma^\mu e \,V_\mu$ & 
$\hat\OO_\chi^\alpha = \frac{4g_\chi g_e}{\Lambda_\chi m_e} \bigl( \delta^{\alpha\beta} - \frac{q^\alpha q^\beta}{q^2} \bigr)\, \hat S_\chi^\beta$ &
$\bar\sigma_e = \frac{g_\chi^2 g_e^2}{\pi} \frac{6m_\chi^2+m_e^2}{\Lambda_\chi^2(m_\chi+m_e)^2}$ \Tstrut\Bstrut\\
\hline
Anapole DM & 
$\LL = \frac{g_\chi}{\Lambda_\chi^2}\bar\chi\gamma^\mu\gamma^5\chi \,\partial^\nu V_{\mu\nu} +g_e\bar e\gamma^\mu e \,V_\mu$ & 
$\hat\OO_\chi^\alpha = \frac{2g_\chi g_e}{\Lambda_\chi^2 m_e} \epsilon^{\alpha\beta\gamma} iq^\beta \hat S_\chi^\gamma$ &
$\bar\sigma_e = \frac{g_\chi^2 g_e^2}{\pi} \frac{3\alpha^2\mu_{\chi e}^2}{2\Lambda_\chi^4}$ \Tstrut\Bstrut\\
\hline
Pseudo-mediated DM & 
$\LL = g_\chi \bar\chi \chi \phi + g_e  \bar e \,i\gamma^5 e \, \phi$ & 
$\hat\OO_\chi^\alpha = -\frac{g_\chi g_e}{q^2 m_e} iq^\alpha \mathbb{1}_\chi$ &
$\bar\sigma_e = \frac{g_\chi^2 g_e^2}{4\pi} \frac{\mu_{\chi e}^2}{\alpha^2 m_e^4}$ \Tstrut\Bstrut\\
\hline
\end{tabular}
\caption{\label{tab:models}
Dark matter models, having Lagrangian ${\cal L}$, with SD interactions considered in this work; these models are particularly well-motivated when DM does not carry a charge of any type, see {\em e.g.} Refs.~\cite{Pospelov:2000bq,Sigurdson:2004zp,Masso:2009mu,Fitzpatrick:2010br,Chang:2010en,Barger:2010gv,Banks:2010eh,Ho:2012bg,Gresham:2014vja,DelNobile:2014eta,Kavanagh:2018xeh,Chu:2018qrm}. $\chi$ is a spin-1/2 DM particle, and $V$, $\phi$ are ultralight (typically $\ll$ eV) spin-1, spin-0 mediators, respectively. $g_\chi$, $g_e$ are dimensionless couplings, and $\Lambda_\chi$ is the effective theory cutoff. In the nonrelativistic limit, these Lagrangians reduce to the operators $\hat\OO_\chi^\alpha$ (with Cartesian coordinates $\alpha=1,2,3$), as in Eq.~\eqref{eq:Ochidef}. $q\equiv|\vect{q}|$ is the momentum transfer, and $\hat S_\chi^\alpha=\sigma^\alpha/2$ is the DM spin operator. $\overline{\sigma}_e$ is the reference cross section defined in Eq.~\eqref{eq:sigmabaredef} that we will use to present the reach.}
\end{table*}

In this {\it Letter}, we propose a novel detection path for spin-dependent light DM-electron interactions via magnon excitations. Magnons are quanta of collective spin wave excitations in condensed matter systems that exhibit magnetic dipole order in the ground state. They can be thought of as the SD counterpart of phonons for DM detection with similar kinematics. We demonstrate as a proof of principle that single magnon excitations can probe interesting DM scenarios through scattering, thus broadening the coverage of the DM theory space.  In future work we will pursue DM (in particular axion DM) absorption through magnon excitations.


\newpar
\paragraph{Magnons in magnetically ordered materials}---\,
Magnetic order can arise in solid state systems due to the interplay between electron-electron interactions, electron kinetic energy and Pauli exclusion (see {\it e.g.}\ Refs.~\cite{auerbach1994interacting,Coey_2001}). Such systems are usually described by a spin lattice model, {\it e.g.} the Heisenberg model,
\begin{equation}
H = \frac{1}{2}\sum_{l,l'=1}^{N} \sum_{j,j'=1}^{n} J_{ll'jj'} \,\vect{S}_{lj}\cdot\vect{S}_{l'j'}\,.
\end{equation}
Here $l,l'$ label the magnetic unit cells, and $j,j'$ label the magnetic atoms/ions inside the unit cell. Depending on the sign of the exchange coupling $J_{ll'jj'}$, the spins $\vect{S}_{lj}$ and $\vect{S}_{l'j'}$ tend to align or anti-align. The low energy excitations are obtained by applying the Holstein-Primakoff transformation to expand the spins around the ordered ground state in terms of bosonic creation and annihilation operators $\hat a^\dagger, \hat a$.  The quadratic part of the Hamiltonian can then be diagonalized via a Bogoliubov transformation (see the Supplemental Material for details),
\begin{align}
\left(\begin{matrix}
\hat a_{j,\vect{k}} \\
\hat a_{j,\vect{-k}}^\dagger
\end{matrix}\right)
&= \left(\begin{matrix}
\mathrm{U}_{j\nu,\vect{k}} & \mathrm{V}_{j\nu,\vect{k}} \\
\mathrm{V}_{j\nu,-\vect{k}}^* & \mathrm{U}_{j\nu,-\vect{k}}^*
\end{matrix}\right)
 \left(\begin{matrix}
\hat b_{\nu,\vect{k}} \\
\hat b_{\nu,\vect{-k}}^\dagger
\end{matrix}\right), \\[4pt]
H &= \sum_{\nu=1}^n\sum_{\vect{k}\in\text{1BZ}} \omega_{\nu,\vect{k}} \hat b_{\nu,\vect{k}}^\dagger \hat b_{\nu,\vect{k}}\,,
\label{eq:Hmagnon}
\end{align}
so that $\hat b^\dagger, \hat b$ are creation and annihilation operators of the canonical magnon modes, which are collective excitations of the spins. For a system with $N$ magnetic unit cells and $n$ magnetic atoms/ions in the unit cell, there are $n$ magnon branches, labeled by $\nu$, with $N$ modes on each branch, labeled by momentum vectors $\vect{k}$ within the first (magnetic) Brillouin zone (1BZ). The $n\times n$ matrices $\mathrm{U}$, $\mathrm{V}$ can be calculated for each $\vect{k}$. 

\newpar
\paragraph{Magnon excitation from dark matter scattering}---\,
If the DM couples to the electron spin, it can scatter off the target material and create magnon excitations.\footnote{Magnons can also be excited via couplings to orbital angular momenta. Here we assume negligible orbital angular momenta for simplicity, noting that this is the case for many familiar materials where 3d electrons are responsible for the magnetic order.
} Suppose the nonrelativistic effective Lagrangian takes the form
\begin{equation}
\LL = -\sum_{\alpha=1}^3\hat\OO_\chi^\alpha(\vect{q}) \hat S_e^\alpha,
\label{eq:Ochidef}
\end{equation}
where $\alpha$ denotes the Cartesian coordinates, and $\vect{q}$ is the momentum transfer from the DM to the target. The operators $\hat\OO_\chi$ that follow from the three Lagrangians we consider are listed in Table~\ref{tab:models}. Focusing on transitions from the ground state to single magnon states $|\nu,\vect{k}\rangle$, we obtain the matrix element as (see the Supplemental Material for details)
\begin{equation}
\MM_{\nu,\vect{k}}^{s_i s_f}(\vect{q}) = \delta_{\vect{q},\vect{k}+\vect{G}}\, \frac{1}{\sqrt{N}\Omega}
\sum_{\alpha=1}^3 \langle s_f| \hat\OO_\chi^\alpha(\vect{q}) |s_i \rangle \,\epsilon_{\nu,\vect{k},\vect{G}}^\alpha\,,
\label{eq:Mmagnon}
\end{equation}
where $\Omega$ is the volume of the magnetic unit cell, $\vect{G}$ denotes a reciprocal lattice vector, and $|s_{i, f} \rangle$ are the initial and final DM spin states. $\boldsymbol{\epsilon}_{\nu,\vect{k},\vect{G}}$ is the analog of polarization vectors for the magnon modes,
\begin{equation}
\boldsymbol{\epsilon}_{\nu,\vect{k},\vect{G}} = \sum_{j=1}^{n} \sqrt{\frac{S_j}{2}} \bigl(\mathrm{V}_{j\nu,-\vect{k}} \vect{r}_j^* +\mathrm{U}_{j\nu,\vect{k}}^* \vect{r}_j \bigr)\, e^{i\vect{G}\cdot\vect{x}_j}\,,
\end{equation}
where $r_j^\alpha\equiv R_j^{\alpha1}+iR_j^{\alpha2}$ parameterize the spin orientations in the ground state,
\begin{equation}
S_{lj}^\alpha = \sum_\beta R_j^{\alpha\beta} S_{lj}'^\beta \,,
\quad
\{\langle S_{lj}'^1\rangle,\, \langle S_{lj}'^2\rangle,\, \langle S_{lj}'^3\rangle\} = \{ 0,\, 0,\, S_j\}\,,
\end{equation}
and $\vect{x}_j \equiv \vect{x}_{lj}-\vect{x}_l$ is the position of the $j$th site within a magnetic unit cell. As a simple example, a ferromagnet with one magnetic ion per unit cell ($n=1$) has $\vect{r}=(1,i,0)$, $U=1$, $V=0$, and thus, $\boldsymbol{\epsilon}=\sqrt{S/2}\,(1,i,0)$ for all $\vect{k}$ and $\vect{G}$, reminiscent of a photon polarization vector.

From Eq.~\eqref{eq:Mmagnon} we see that for given $\vect{q}$, only the magnon modes with $\vect{k}\in\text{1BZ}$ satisfying $\vect{q}=\vect{k}+\vect{G}$ for some $\vect{G}$ can be excited, due to lattice momentum conservation. Summing over $s_f$ and averaging over $s_i$, we obtain
\begin{equation}
\overline{|\MM_{\nu,\vect{k}}(\vect{q})|^2}
= \frac{\delta_{\vect{q},\vect{k}+\vect{G}}}{N\Omega^2} 
\,\text{tr} \bigl( \hat\rho_\chi\hat\OO_\chi^\alpha(\vect{q}) \hat\OO_\chi^{\dagger\beta}(\vect{q}) \bigr)\,\epsilon_{\nu,\vect{k},\vect{G}}^\alpha \epsilon_{\nu,\vect{k},\vect{G}}^{*\beta}\,,
\end{equation}
where $\hat\rho_\chi =\frac{1}{2S_\chi+1} \mathbb{1}_{2S_\chi+1}$ is the density matrix for the spin of the incoming DM. The total event rate per unit target mass is then obtained as
\begin{align}
R &= \frac{1}{\rho_T} \frac{\rho_\chi}{m_\chi}\int d^3v_\chi \,f(\vect{v}_\chi) \sum_\nu\sum_{\vect{k}\in\text{1BZ}}\Gamma_{\nu,\vect{k}}(\vect{v}_\chi)\,,\\
\Gamma_{\nu,\vect{k}}(\vect{v}_\chi) &= 2\pi\sum_{\vect{q}=\vect{k}+\vect{G}} \overline{|\MM_{\nu,\vect{k}}(\vect{q})|^2} \,\delta\left(E_{\chi_i}-E_{\chi_f}-\omega_{\nu,\vect{k}}\right)\,,
\end{align}
where $\rho_T$ is the target mass density, $\rho_\chi=0.3\,\text{GeV}/\text{cm}^3$ is the local DM energy density, $E_{\chi_i}=\frac{1}{2}m_\chi v_\chi^2$, $E_{\chi_f}=(m_\chi \vect{v}_\chi-\vect{q})^2/(2 m_\chi)$. We assume the DM velocity distribution $f(\vect{v}_\chi)$ is Maxwell-Boltzmann, with dispersion 220\,km/s, truncated by the galactic escape velocity 500\,km/s, and boosted to the target rest frame by the Earth's velocity in the galactic rest frame, 240\,km/s. We take the continuum limit $\sum_{\vect{k}\in\text{1BZ}}\to N\Omega\int\frac{d^3k}{(2\pi)^3}$, where $R$ becomes $N$ independent.

\newpar
\paragraph{Projected reach}---\,
As a first demonstration of the detection concept, we consider a yttrium iron garnet (YIG, Y$_3$Fe$_5$O$_{12}$) target. YIG is a classic ferrimagnetic material that has been extensively studied and well-characterized, and can be readily synthesized with high quality \cite{Saga,Princep_2017}. It has been exploited for axion DM detection via absorption in an external magnetic field~\cite{Barbieri:1985cp,Crescini:2018qrz,Flower:2018qgb}. Here we focus on DM scattering for which external fields are not necessary for producing a signal. Particular detection schemes will be explored in future work.

YIG has 20 magnetic ions Fe$^{3+}$ per unit cell, with effective spins $S_j=5/2$ ($j=1,\dots,20$) coming from five 3d electrons with quenched orbital angular momentum. The ground state has the 12 tetrahedral-site and 8 octahedral-site spins pointing in opposite directions. Taking the crystal parameters from Ref.~\cite{MaterialsProject} and Heisenberg model parameters from Ref.~\cite{Saga}, we diagonalize the magnon Hamiltonian using the algorithm of Ref.~\cite{Toth-Lake} to obtain the magnon spectrum $\omega_{\nu,\vect{k}}$ and the $\mathrm{U},\mathrm{V}$ matrices that enter the rate formulae. For simplicity, we fix the direction of the DM wind to be parallel (perpendicular) to the ground state spins for the magnetic dipole and anapole (pseudo-mediated) models, which maximizes the event rate. For fixed target orientation, we find a daily modulation of $\OO(10\%)$, which could be utilized for distinguishing DM signals from backgrounds. Following common practice, we present the projected reach in terms of a reference cross section $\bar\sigma_e$ defined from DM-free electron scattering. Here we generalize the definition in Ref.~\cite{Essig:2011nj} beyond SI interactions by defining
\begin{equation}
\bar\sigma_e \equiv \frac{\mu_{\chi e}^2}{16\pi m_\chi^2 m_e^2} \overline{|\MM_\text{free}|^2}\,(q=\alpha m_e, v^\perp=\alpha)\,,
\label{eq:sigmabaredef}
\end{equation}
where $\mu_{\chi e}$ is the DM-electron reduced mass, $\alpha=1/137$ is the fine structure constant, and $v^\perp$ is the component of the relative velocity perpendicular to $\vect{q}$.  The reference cross section for each model is given in Table~\ref{tab:models}.

\begin{figure*}[t]
\includegraphics[width=\linewidth]{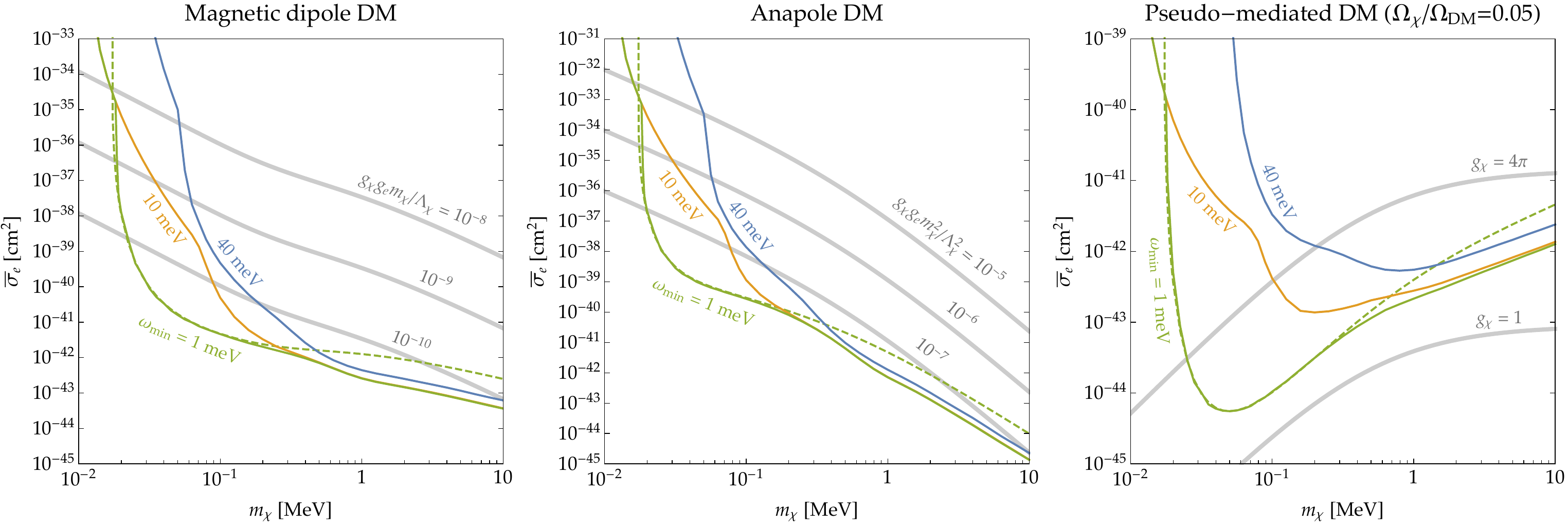}
\caption{\label{fig:reach}
Projected reach for the DM models in Table~\ref{tab:models} for a YIG target, assuming three events with kilogram-year exposure, for several magnon detection thresholds $\omega_\text{min}$ (solid). Also shown are the results of a Heisenberg ferromagnet with the same mass and spin densities as YIG, and the same magnon dispersion as the low-energy gapless modes of YIG, for $\omega_\text{min}=1\,$meV (dashed); they coincide with the YIG curves for $0.02\,\text{MeV}\lesssim m_\chi\lesssim0.1\,\text{MeV}$, which can be understood from the effective theory argument in the text. The gray contours show the model parameters in the magnon sensitivity regions, which astrophysical and cosmological constraints on specified UV completions can be mapped onto (see text). For the pseudo-mediated model, we consider a DM subcomponent to evade SIDM constraints, and let $g_e$ saturate the white dwarf cooling bound.
}
\end{figure*}

Our results are shown in Fig.~\ref{fig:reach} for $m_\chi$ up to 10\,MeV, assuming 3 events on a YIG target (colored solid curves) with kilogram-year exposure and, following convention for easy comparison to other experiments, no background.\footnote{For calorimetric readout, the backgrounds are expected to be similar to other experiments reading out meV-eV energy depositions: radiogenic backgrounds are not expected to be problematic at such low energies, while coherent scattering from high-energy photons can be suppressed with an active veto, leaving $pp$ solar neutrinos the main irreducible background. We expect the latter to be at most a few events per kilogram-year, as estimated from neutrino-nucleus scattering (see {\it e.g.}\ Refs.~\cite{Hochberg:2015fth,Essig:2018tss}).} Beyond 10\,MeV, the simple Heisenberg model description breaks down in part of the kinematic integration region where $q$ exceeds the inverse ionic radius of Fe$^{3+}$; however, electron excitations are expected to have sensitivity in this mass regime \cite{Essig:2011nj,Graham:2012su,Essig:2012yx,Lee:2015qva,Essig:2015cda,Derenzo:2016fse,Essig:2017kqs,Kurinsky:2019pgb} (though precise results are not currently available for the SD models considered here). We consider several detector thresholds $\omega_\text{min}$ corresponding to capabilities of TESs expected within the next few years (40\,meV) and further into the future (10\,meV, 1\,meV). Also shown in the plots are contours of model parameters in the magnon sensitivity region (gray).

For each benchmark DM model, magnons can probe currently unconstrained parameter space. For the vector mediator models, assuming the mediator $V$ couples to SM particles only via kinetic mixing with the photon, $V$ production in stellar media and in the early universe is suppressed when $m_V\to0$, so the only astrophysical and cosmological constraints are from DM production. The latter, however, depend on whether $\Lambda_\chi$ is above or below the energies involved and, if below, the ultraviolet (UV) completion of the effective operators. For example, if $\Lambda_\chi\sim m_\chi$ and the UV completion involves millicharged particles \cite{Davidson:2000hf,Vogel:2013raa} with couplings $\sim g_\chi$, we find that magnetic dipole DM with $g_\chi g_e\lesssim 10^{-10}$ satisfies all existing constraints, but can be probed by magnons. On the other hand, if $\Lambda_\chi\gtrsim\OO(100\,\text{MeV})$, we can map the constraints derived in Ref.~\cite{Chu:2019rok} onto the gray contours in Fig.~\ref{fig:reach}, e.g.\ excluding $g_\chi g_e m_\chi /\Lambda_\chi \in (10^{-12}, \,10^{-10}) \,m_\chi/m_e$ for magnetic dipole DM from SN1987A --- in this case, there is a large region of unconstrained parameter space above this band (even after imposing Big Bang Nucleosynthesis constraints)~\cite{Chu:2019rok}, which can be fully covered by our projected magnon reach. The anapole model is more challenging to discover via magnons due to the high power of momentum suppression, but the magnon sensitivity region still accommodates viable UV models, such as those involving two dark photons~\cite{Masso:2006gc} which evade astrophysical and cosmological bounds altogether. Finally, for the pseudo-mediated DM model, the mediator-electron coupling is constrained by white dwarf cooling to be $g_e\lesssim 2\times 10^{-13}$, so that $g_\chi$ has to be $\OO(1)$ to produce a detectable signal. Given the existing self-interacting dark matter constraints, we consider $\chi$ to be a 5\% subcomponent of DM as a viable scenario, and show contours of $g_\chi$ in Fig.~\ref{fig:reach} with $g_e$ saturating its upper bound.

To gain some analytical intuition, we note that for momentum transfer well within the 1BZ, corresponding to $m_\chi\lesssim 0.1\,$MeV for a YIG target, the rate can be estimated via an effective $n=1$ ferromagnetic model. This is because in the $q\to0$ limit, the external probe $\hat\OO_\chi^\alpha$ acts like a uniform magnetic field. In a semiclassical picture, this causes all the spins in the target to precess in phase, so the angle between them, and thus the total energy of the Heisenberg model, stays the same. As a result, only the {\it gapless} mode(s), {\it i.e.}\ Goldstone mode(s) of the broken rotational symmetry, can be excited. Even for finite $q$, gapped magnon contributions are suppressed by powers of $aq$, where $a$ is the lattice spacing, and thus subdominant for $q\ll a^{-1}$ ($\simeq 0.2\,$keV for YIG). For a ferrimagnet like YIG, we can integrate out the gapped modes to arrive at an effective theory, where the only relevant degree of freedom is the total spin density $n_s$. There is only one magnon branch in this effective $n=1$ ferromagnetic theory, which matches the gapless branch of the original ferrimagnet for $k\ll a^{-1}$. For YIG, the total spin density is $S_\text{cell}=(12-8)\times 5/2 = 10$ per unit cell volume $\Omega=a^3/2$, with $a\simeq 12.56\,\text{\AA}$, {\it i.e.}\ $n_s=20/a^3\simeq (4.6\,\text{\AA})^{-3}$. The effective exchange coupling can be shown to be $J_\text{eff}\simeq -4\,\text{K}=-0.35\,\text{meV}$ \cite{Saga}, resulting in a quadratic magnon dispersion $\omega=|J_\text{eff}|S_\text{cell}\,(ak)^2\simeq k^2/(7\,\text{MeV})$ at small $k$. For this $n=1$ ferromagnetic theory, we obtain (see the Supplemental Material for details),
\begin{align}
R \simeq& \frac{n_s}{\rho_T} \frac{\rho_\chi}{m_\chi}\int d^3v_\chi \,f(\vect{v}_\chi)\,\cdot \nonumber\\
&\quad \int\frac{d^3q}{8\pi^2}\,\text{tr} \bigl( \hat\rho_\chi\hat\OO_\chi^+(\vect{q}) \hat\OO_\chi^{\dagger-}(\vect{q}) \bigr)
\,\delta\left(E_{\chi_i}-E_{\chi_f}-\omega\right)\,,\nonumber\\
\simeq& 3\, (\text{kg}\cdot\text{yr})^{-1}\, \biggl(\frac{n_s}{(4.6\,\text{\AA})^{-3}}\biggr) \biggl(\frac{4.95\,\text{g}/\text{cm}^3}{\rho_T}\biggr) \biggl(\frac{0.1\,\text{MeV}}{m_\chi}\biggr) \nonumber\\
&\quad \int d^3v_\chi \,f(\vect{v}_\chi) \biggl(\frac{10^{-3}}{v_\chi}\biggr) \biggl(\frac{\hat R}{4\times10^{-27}}\biggr)\,,
\label{eq:Ranalytical}
\end{align}
where $\hat\OO_\chi^\pm\equiv \hat\OO_\chi^1\pm i\hat\OO_\chi^2$, and
\begin{align}
\hat R &= m_e^2\int\frac{d^3q}{2\pi q} \,\text{tr}\bigl(\hat\rho_\chi\hat\OO_\chi^+\hat\OO_\chi^{\dagger-}\bigr)\,\delta\biggl(\cos\theta-\frac{q}{2m_\chi v_\chi} -\frac{\omega}{v_\chi q}\biggr) \nonumber\\[4pt]
&=
\begin{cases}
\frac{2g_\chi^2 g_e^2(1+\langle c^2\rangle)}{\Lambda_\chi^2} (q_\text{max}^2-q_\text{min}^2) & \text{(magnetic dipole)}\,, \\[4pt]
\frac{g_\chi^2 g_e^2(1+\langle c^2\rangle)}{4 \Lambda_\chi^4} (q_\text{max}^4-q_\text{min}^4) & \text{(anapole)}\,, \\[4pt]
g_\chi^2 g_e^2 \langle s^2\rangle\log(q_\text{max}/q_\text{min}) & \text{(pseudo-mediated)}\,.
\end{cases}
\label{eq:Rhat}
\end{align}
Here $\theta$ is the angle between $\vect{q}$ and $\vect{v}_\chi$, $\langle c^2\rangle$ and $\langle s^2\rangle$ are properly averaged values of cosine and sine squared of the angle between $\vect{q}$ and the ground state spin direction over accessible scattering kinematics, $q_\text{max}\simeq 2m_\chi v_\chi$, and $q_\text{min}$ is the magnon momentum for which $\omega_{\vect{q}}=\omega_\text{min}$. The $q$ dependence in Eq.~\eqref{eq:Rhat} is indicative of dipole-dipole, quadrupole-dipole and charge-dipole type interactions, respectively, for the three DM models.

The projected reach for this $n=1$ Heisenberg ferromagnet is shown by the dashed curves in Fig.~\ref{fig:reach} in the $\omega_\text{min}=1$\,meV case, with $\langle c^2\rangle$ set to $1/3$. We see that the full YIG results are almost exactly reproduced for $0.02\,\text{MeV}\lesssim m_\chi \lesssim 0.1\,\text{MeV}$. For $m_\chi\lesssim 0.02\,\text{MeV}$, the gapless branch becomes kinematically inaccessible, and the reach is dominated by the gapped magnons. For $m_\chi\gtrsim 0.1\,\text{MeV}$, YIG beats the $n=1$ ferromagnet due to contributions from the gapped magnons, which are no longer suppressed as the typical momentum transfer approaches (and goes beyond) the boundaries of the 1BZ. For higher $\omega_\text{min}$, effective theory predictions (not shown) are off because the lowest-energy magnon modes on the gapless branch become inaccessible.

\newpar
\paragraph{Discussion}---\,
While we have chosen three specific DM models for illustration, we note that there are other scenarios with SD interactions that can be probed via magnon excitation. Examples include models with a spin-1 mediator coupling to $\bar e \gamma^\mu\gamma^5 e$ or nonminimally to the electron. Generally, $\hat\OO_\chi^\alpha$ is the mediator propagator multiplied by a function that is at least linear in $q$, so the rate is at least logarithmic (as in the pseudo-mediated model). Given the strong astrophysical and cosmological constraints on light DM and mediator scenarios \cite{Hochberg:2015fth,Green:2017ybv,Knapen:2017xzo}, magnon excitations are most relevant for probing subcomponents of DM with SD interactions, if not mediated by a dark photon.

Beyond scattering, a magnon signal can also arise from absorption of bosonic DM. A prime example is an axion $a$ interacting via $(\partial_\mu a)\,\bar e \gamma^\mu\gamma^5 e\to \nabla a\cdot\vect{S}_e$. However, Heisenberg-type materials with 3d electrons, such as YIG, have very limited sensitivity to DM absorption, because gapped modes with $k\simeq 0$, which match the kinematics, have strongly suppressed matrix elements as explained above. Here we identify three possible solutions to pursue in future work. First, in materials with nondegenerate Land\'e $g$-factors (due to different orbital angular momentum admixtures in the effective spins), magnetic atoms/ions within the same unit cell can respond differently in the $q\to0$ limit, allowing excitation of gapped magnons. Second, anisotropic spin-spin interactions can lift the otherwise gapless Goldstone modes, enabling them to match DM absorption kinematics. Finally, the gapless modes can also be lifted by an external magnetic field, which can be tuned to scan the DM mass, as considered in Refs.~\cite{Barbieri:1985cp,Crescini:2018qrz,Flower:2018qgb} (see also Ref.~\cite{Marsh:2018dlj}) in the context of axion absorption.

\newpar
\paragraph{Conclusions}---\,
Collective excitations in condensed matter systems offer a novel detection path for light DM because of favorable kinematics. Given our ignorance of how the DM may interact with SM particles, it is important to explore different types of collective excitations in various materials in order to cover the broadest range of possibilities. In this {\it Letter}, we proposed using magnon excitations to detect DM in the $10$\,keV-$10$\,MeV mass range that couples to the electron spin. This complements previous proposals of detecting spin-independent DM interactions via phonon excitation. For a concrete demonstration of the discovery potential, we calculated the rate for three benchmark DM models, and found that currently unconstrained parameter space can be probed via magnon excitation in a YIG target.

To move forward and realize our proposed DM detection concept, a pressing question is an experimental scheme to detect magnon quanta. One possibility is calorimetric readout similar to phonon detection~\cite{Knapen:2017ekk,Griffin:2018bjn}, in which case magnon propagation and decay, as well as magnon-TES/MKID interactions, need to be understood. Besides, recent research in quantum magnonics has taken on the challenge of resolving single magnons~\cite{Lachance_Quirion_2017,Lachance-Quirion:aa}, and may find application in DM detection. We plan to investigate these possibilities in future work.


\newpar
\paragraph{Acknowledgments}---\,
We thank Sin\'ead Griffin, David Hsieh, Matt Pyle and Mengxing Ye for useful discussions, and Sin\'ead Griffin, Katherine Inzani and Thomas Harrelson for collaboration on related work. T.T., Z.Z., and K.Z.\ were supported by the DOE under Contract No.~DE-AC02-05CH11231, and by the Quantum Information Science Enabled Discovery (QuantISED) for High Energy Physics (Grant No.~KA2401032). Z.Z.\ was also supported by the NSF Grant No.~PHY-1638509. We thank the CERN theory group for hospitality where part of this work was completed. Z.Z.\ and K.Z.\ also thank the Aspen Center for Physics for hospitality during the completion of this work.

\bibliography{magnon}

\onecolumngrid
\vskip40pt
\section{Supplemental material:\\additional details of the magnon rate calculation}

\renewcommand{\theequation}{A.\arabic{equation}}
\setcounter{equation}{0}

\begin{figure}[t]
\includegraphics[width=0.5\linewidth]{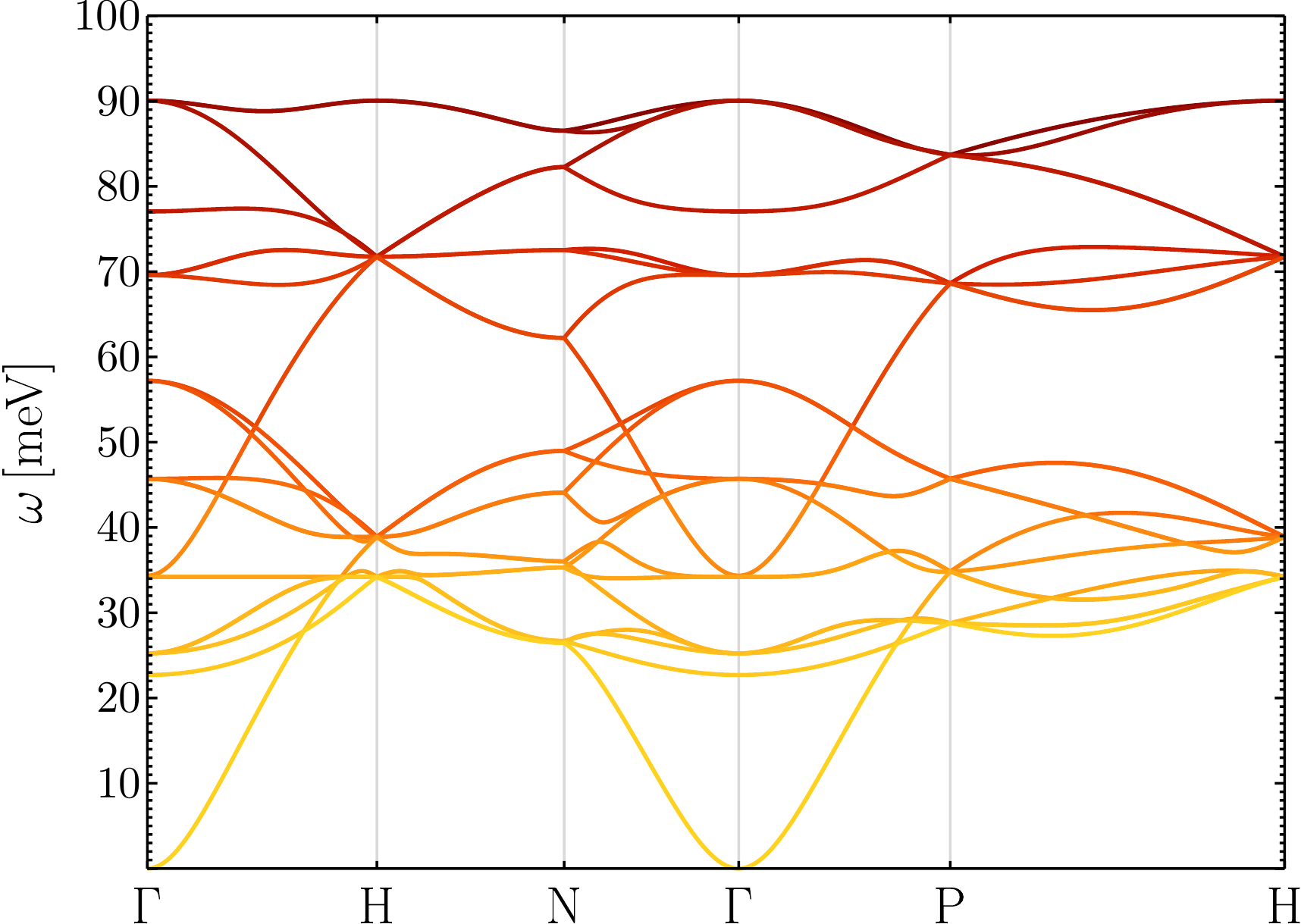}
\caption{\label{fig:dispersion}
Calculated magnon dispersion of YIG along the high symmetry lines in the first Brillouin zone.}
\end{figure}

Here we provide additional technical details of the calculations in the {\it Letter}. We begin by reviewing the derivation of the magnon Hamiltonian (Eq.~(3) in the {\it Letter}). We first define a local coordinate system for each sublattice $j$, in which the spins point in the $z$ direction in the ground state. Denoting the rotation matrices between global and local coordinates by $R_j$, we have
\begin{equation}
S_{lj}^\alpha = \sum_\beta R_j^{\alpha\beta} S_{lj}'^\beta \,,
\quad
\{\langle S_{lj}'^1\rangle,\, \langle S_{lj}'^2\rangle,\, \langle S_{lj}'^3\rangle\} = \{ 0,\, 0,\, S_j\}\,,
\label{eq:Rdef}
\end{equation}
where $\alpha$, $\beta$ are Cartesian coordinates. To find the excitations above the ground state, we map the spin system onto a bosonic system via the Holstein-Primakoff transformation,
\begin{equation}
S_{lj}'^+ = \bigl(2S_j-\hat a_{lj}^\dagger \hat a_{lj}\bigr)^{1/2}\, \hat a_{lj}\,,\qquad
S_{lj}'^- = \hat a_{lj}^\dagger\,\bigl(2S_j-\hat a_{lj}^\dagger \hat a_{lj}\bigr)^{1/2}\,,\qquad
S_{lj}'^3 = S_j -\hat a_{lj}^\dagger \hat a_{lj}\,,
\end{equation}
where $S_{lj}'^\pm=S_{lj}'^1\pm iS_{lj}'^2$. The bosonic creation and annihilation operators satisfy $[\hat a_{lj}, \hat a_{l'j'}^\dagger]=\delta_{ll'}\delta_{jj'}$, so that commutators between the spin operators $[S_{lj}'^\alpha, S_{l'j'}'^\beta] = \delta_{ll'}\delta_{jj'}\,i\epsilon^{\alpha\beta\gamma}S_{lj}'^\gamma$ are reproduced. Going to momentum space and diagonalizing the quadratic Hamiltonian (corresponding to the leading terms in the $1/S$ expansion) by a Bogoliubov transformation, 
\begin{eqnarray}
\hat a_{lj} &=& \frac{1}{\sqrt{N}} \sum_{\vect{k}\in\text{1BZ}} \hat a_{j,\vect{k}} e^{i\vect{k}\cdot\vect{x}_{lj}}\,,
\label{eq:Fourier}\\[2pt]
\left(\begin{matrix}
\hat a_{j,\vect{k}} \\
\hat a_{j,\vect{-k}}^\dagger
\end{matrix}\right)
&=& \mathrm{T}_{\vect{k}}
 \left(\begin{matrix}
\hat b_{\nu,\vect{k}} \\
\hat b_{\nu,\vect{-k}}^\dagger
\end{matrix}\right) 
\qquad\text{where}\quad
\mathrm{T}_{\vect{k}} = 
\left(\begin{matrix}
\mathrm{U}_{j\nu,\vect{k}} & \mathrm{V}_{j\nu,\vect{k}} \\
\mathrm{V}_{j\nu,-\vect{k}}^* & \mathrm{U}_{j\nu,-\vect{k}}^*
\end{matrix}\right),
\label{eq:diag}
\end{eqnarray}
where $\vect{x}_{lj}$ is the position of the $j$th site in the $l$th unit cell, we arrive at the free magnon Hamiltonian,
\begin{equation}
H = \sum_{\nu=1}^n\sum_{\vect{k}\in\text{1BZ}} \omega_{\nu,\vect{k}} \hat b_{\nu,\vect{k}}^\dagger \hat b_{\nu,\vect{k}}\,,
\end{equation}
where $b_{\nu,\vect{k}}^\dagger, \hat b_{\nu,\vect{k}}$ are creation and annihilation operators for the canonical magnon modes. The canonical commutators are preserved, $[\hat b_{\nu,\vect{k}}, \hat b_{\nu',\vect{k}'}^\dagger]=\delta_{\nu\nu'}\delta_{\vect{k}\vect{k}'}$, by imposing the following constraint,
\begin{equation}
\mathrm{T}_{\vect{k}}
\left(\begin{matrix}
\mathbb{1}_n & \mathbb{0}_n \\
\mathbb{0}_n & -\mathbb{1}_n
\end{matrix}\right)
\mathrm{T}_{\vect{k}}^\dagger =
\left(\begin{matrix}
\mathbb{1}_n & \mathbb{0}_n \\
\mathbb{0}_n & -\mathbb{1}_n
\end{matrix}\right).
\end{equation}
We follow the algorithm in Ref.~\cite{Toth-Lake} to solve the constrained diagonalization problem to obtain $\omega_{\nu,\vect{k}}$, $\mathrm{T}_{\vect{k}}$. Note that Ref.~\cite{Toth-Lake} uses a different Fourier transformation convention, with $\vect{x}_l$ rather than $\vect{x}_{lj}$ in the exponent of Eq.~\eqref{eq:Fourier}. We have consistently followed our convention throughout the calculation, adjusting the equations in Ref.~\cite{Toth-Lake} where necessary. In Fig.~\ref{fig:dispersion}, we plot our calculated magnon dispersion $\omega_{\nu,\vect{k}}$ for YIG along the high symmetry lines in the (body-centered cubic) 1BZ generated using the SeeK-path code \cite{Hinuma:aa}.

Next, we derive the single magnon production matrix element (Eq.~(5) in the {\it Letter}) from the DM-electron spin coupling (Eq.~(4) in the {\it Letter}). Assuming the absence of orbital angular momentum, a magnetic atom/ion at site $l,j$ sources an effective scattering potential for the incoming DM, which is given by the Fourier transform of the momentum space operator,
\begin{equation}
V_{lj}(\vect{x}) = \int \frac{d^3q}{(2\pi)^3} \sum_\alpha\hat\OO_\chi^\alpha(\vect{q}) \hat S_{lj}^\alpha\, e^{-i\vect{q}\cdot(\vect{x}-\vect{x}_{lj})}\,.
\end{equation}
For a DM particle with incoming momentum $\vect{p}$ and outgoing momentum $\vect{p}'=\vect{p}-\vect{q}$, and a transition $\lambda_i\to\lambda_f$ in the target system, the matrix element is
\begin{equation}
\MM = \langle \chi_f\lambda_f| \hat V |\chi_i\lambda_i\rangle
= \frac{1}{N\Omega} \sum_{lj} \int d^3x \,e^{i\vect{q}\cdot\vect{x}} \langle s_f\lambda_f| V_{lj} (\vect{x}) | s_i \lambda_i\rangle
= \frac{1}{N\Omega}\sum_\alpha \langle s_f|\hat\OO_\chi^\alpha(\vect{q})| s_i\rangle \sum_{lj} e^{i\vect{q}\cdot\vect{x}_{lj}} \langle\lambda_f |\hat S_{lj}^\alpha| \lambda_i\rangle\,.
\label{eq:Mmagnoncalc1}
\end{equation}
Now focus on the case where $\lambda_i$ is the ground state $|0\rangle$ and $\lambda_f$ is a single magnon state $|\nu,\vect{k}\rangle$. Plugging in Eqs.~\eqref{eq:Rdef}-\eqref{eq:diag}, and keeping only terms proportional to a single power of $\hat b^\dagger_{\nu,\vect{k}}$, we obtain
\begin{eqnarray}
\sum_{lj} e^{i\vect{q}\cdot\vect{x}_{lj}} \langle\lambda_f |\hat S_{lj}^\alpha| \lambda_i\rangle &=&  \sum_{lj} e^{i\vect{q}\cdot\vect{x}_{lj}} \sum_\beta \langle \nu,\vect{k}|R_j^{\alpha\beta}\hat S_{lj}'^\beta| 0\rangle 
= \sum_{lj} \sqrt{\frac{S_j}{2}}  e^{i\vect{q}\cdot\vect{x}_{lj}} \langle \nu,\vect{k}|r_j^{\alpha*}\hat a_{lj}+r_j^\alpha \hat a_{lj}^\dagger| 0\rangle \nonumber\\
&=& \frac{1}{\sqrt{N}} \sum_{\vect{k}'\in\text{1BZ}} \sum_l  e^{i(\vect{q}-\vect{k}')\cdot\vect{x}_l}  \sum_j\sqrt{\frac{S_j}{2}} \,e^{i(\vect{q}-\vect{k}')\cdot\vect{x}_j} \langle \nu,\vect{k}|r_j^{\alpha*}\hat a_{j,-\vect{k}'}+r_j^\alpha \hat a_{j,\vect{k}'}^\dagger| 0\rangle \nonumber\\
&=& \sqrt{N} \sum_{\vect{k}'\in\text{1BZ}} \sum_{\vect{G}} \delta_{\vect{q}-\vect{k}',\vect{G}} \sum_{j\nu'}\sqrt{\frac{S_j}{2}} \,e^{i\vect{G}\cdot\vect{x}_j} \bigl(\mathrm{V}_{j\nu',-\vect{k}'}r_j^{\alpha*} +\mathrm{U}^*_{j\nu',\vect{k}'}r_j^\alpha \bigr)\,\langle \nu,\vect{k}|\hat b_{\nu',\vect{k}'}^\dagger| 0\rangle \nonumber\\
&=& \sqrt{N} \sum_{\vect{G}} \delta_{\vect{q}-\vect{k},\vect{G}} \sum_j \sqrt{\frac{S_j}{2}} \,e^{i\vect{G}\cdot\vect{x}_j} \bigl(\mathrm{V}_{j\nu,-\vect{k}}r_j^{\alpha*} +\mathrm{U}^*_{j\nu,\vect{k}}r_j^\alpha \bigr)\,,
\label{eq:Mmagnoncalc2}
\end{eqnarray}
where we have used $\vect{x}_{lj}=\vect{x}_l+\vect{x}_j$, $\sum_l e^{i(\vect{q}-\vect{k}')\cdot\vect{x}_l}=N\sum_{\vect{G}} \delta_{\vect{q}-\vect{k}',\vect{G}}$. Plugging Eq.~\eqref{eq:Mmagnoncalc2} into Eq.~\eqref{eq:Mmagnoncalc1} reproduces Eq.~(5) in the {\it Letter} (where the sum over $\vect{G}$ is implicit).

Finally, we derive the analytical approximation for the rate in the case of an $n=1$ ferromagnet target (Eqs.~(12) and (13) in the {\it Letter}). Noting that the $\vect{k}$ integral over the 1BZ combined with the $\vect{G}$ sum is equivalent to an integral over the entire momentum space, we have
\begin{equation}
R = \frac{n_s}{\rho_T} \frac{\rho_\chi}{m_\chi}\int d^3v_\chi \,f(\vect{v}_\chi) \int\frac{d^3q}{8\pi^2} \,\text{tr} \bigl( \hat\rho_\chi\hat\OO_\chi^+(\vect{q}) \hat\OO_\chi^{\dagger-}(\vect{q}) \bigr) \,\delta\Bigl(\vect{q}\cdot\vect{v}_\chi -\frac{q^2}{2m_\chi} -\omega_{\vect{k}=\vect{q}-\vect{G}}\Bigr).
\end{equation}
Since the magnon dispersion is near isotropic, the delta function fixes the angle between $\vect{q}$ and $\vect{v}_\chi$ for any given $q=|\vect{q}|$ --- this is true as long as $\frac{q^2}{2m_\chi}+\omega \le qv_\chi$, or approximately (since $\omega\ll qv_\chi$), $q\le 2m_\chi v_\chi$. Thus,
\begin{equation}
R = \frac{n_s}{\rho_T} \frac{\rho_\chi}{m_\chi}\int d^3v_\chi \,\frac{f(\vect{v}_\chi)}{v_\chi} \int_\Sigma\frac{dq\, d\phi}{8\pi^2}\, q \,\text{tr} \bigl( \hat\rho_\chi\hat\OO_\chi^+(\vect{q}) \hat\OO_\chi^{\dagger-}(\vect{q}) \bigr),
\end{equation}
where the $q$ integral is now over a two-dimensional surface $\Sigma$ that satisfies the energy-conserving delta function, which is approximately a sphere of radius $m_\chi v_\chi$ centered at $m_\chi\vect{v}_\chi$. The trace generally depends on the angle $\theta_q$ between $\vect{q}$ and the spins. For the three models in Table~\ref{tab:models}, we have
\begin{equation}
\text{tr}\bigl(\hat\rho_\chi\hat\OO_\chi^+\hat\OO_\chi^{\dagger-}\bigr)
=
\begin{cases}
\frac{4g_\chi^2 g_e^2}{\Lambda_\chi^2 m_e^2} \,(1+\cos^2\theta_q)  & \text{(magnetic dipole DM)}\,, \\[4pt]
\frac{g_\chi^2 g_e^2}{\Lambda_\chi^4 m_e^2} \,q^2 \,(1+\cos^2\theta_q) & \text{(anapole DM)}\,, \\[4pt]
\frac{y_\chi^2 y_e^2}{m_e^2} \,\frac{1}{q^2}\,\sin^2\theta_q & \text{(pseudo-mediated DM)}\,.
\end{cases}
\end{equation}
However, since the trigonometric functions are bounded, we have, {\it e.g.}, $\int dq d\phi \,f(q) \cos^2\theta_q = 2\pi\langle c^2\rangle \int dq\, f(q)$ for a general function $f(q)$, with $\langle c^2\rangle\in [0,1]$ a constant to be understood as a weighted average of $\cos^2\theta_q$ over the integration region. Thus,
\begin{equation}
R = \frac{n_s}{\rho_T} \frac{\rho_\chi}{m_\chi}\int d^3v_\chi \,\frac{f(\vect{v}_\chi)}{v_\chi} \int_{q_\text{min}(\omega_\text{min})}^{q_\text{max}(v_\chi)} \frac{dq}{4\pi} \,q \,\bigl\langle\text{tr} \bigl( \hat\rho_\chi\hat\OO_\chi^+(\vect{q}) \hat\OO_\chi^{\dagger-}(\vect{q}) \bigr)\bigr\rangle.
\end{equation}
After performing the $q$ integral, we arrive at Eqs.~(12) and (13) in the {\it Letter}.

\end{document}